# SNS Standard Power Supply Interface

S. Peng, R. Lambiase, B. Oerter, J. Smith BNL, Upton, NY 11973, USA

## Abstract

The SNS has developed a standard power supply interface for the approximately 350 magnet power supplies in the SNS accumulator ring, Linac and transport lines. Power supply manufacturers are providing supplies compatible with the standard interface. The SNS standard consists of a VME based power supply controller module (PSC) and a power supply interface unit (PSI) that mounts on the power supply. Communication between the two is via a pair of multimode fibers. This PSI/PSC system supports one 16-bit analog reference, four 16-bit analog readbacks, fifteen digital commands and sixteen digital status bits in a single fiber-isolated module. The system can send commands to the supplies and read data from them synchronized to an external signal at up to a 10KHz rate. The PSC time stamps and stores this data in a circular buffer so historical data leading up to a fault event can be analyzed. The PSC contains a serial port so that local testing of hardware can be accomplished with a laptop. This paper concentrates on the software being provided to control the power supply. It includes the EPICS driver; software to test hardware and power supplies via the serial port and VME interface.

# 1 HARDWARE

## 1.1 Block Diagram

A block diagram of the hardware and the connection to the power supplies and control system is given in Figure 1

## 1.2 PSI

The PSI is directly connected to the power supply with a pair of short cables and to the PSI with a pair of fibers. The PSI has four 16-bit ADCs, one 16-bit DAC output, 15 command output bits and accepts 16 status bits.

The PSI is controlled by commands coming from the PSC over fiber cables. PSC commands are used to send new DAC values, send commands (on/off), read the analog input signals and status bits, read the last command or setpoint.

The SNS power supplies require pulses for On, Off and Standby commands. For versatility the PSI has a jumper to give the option of having pulsed or static commands on some command lines. The PSI checks the message received for the correct checksum. If the checksum is wrong, the message is returned to the PSC.

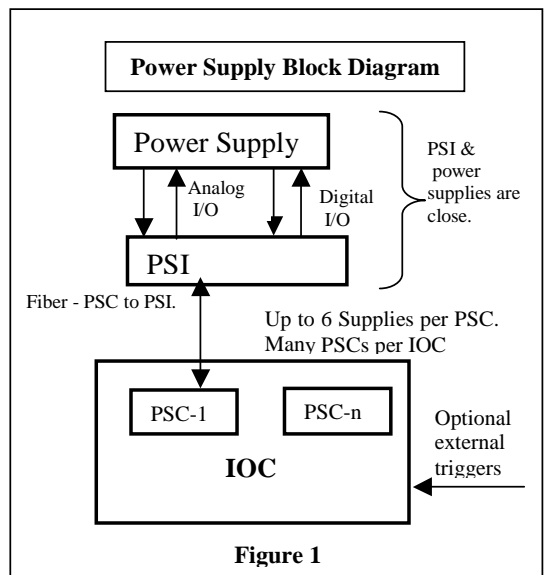

**Figure 1**

The ADC's in the PSI are issued a calibrate command when powered on. There is a delay of 60 seconds to ensure that all voltages have settled. A recalibrate command can be issued at any time from the PSC.

## 1.3 PSC

The PSC is a VME board that communicates with up to 6 PSIs controllers via fibers. The PSC and PSI can be separated by up to 4000 meters. The PSC is designed to operate with software initiated or external triggers. If setup for external triggers, each time a pulse is received a command is sent to the PSI. The time from trigger start until command received at the PSI is fixed, less than 100 microseconds. There are separate triggers for read and write commands. . By adjusting the external trigger timing the user can decide exactly what time the ADCs are read or the DAC is written. Timed reads are useful in those applications where it is necessary to sample data at a particular time. As an example in the SNS we want to know if the kicker magnets are fully charged just before injection starts. With hardware-triggered reads, data can be read at a high rate and the data saved in PSC memory. This is done without any CPU overhead. By reading at a high rate the software can optionally average data to get more accurate readings.

When the status and ADC data is read from the PSI the data is stored in a circular buffer. Each channel has a separate buffer. The buffer size allows saving 5458 frames of data. A frame contains four 16-bit analog input values plus 16bits of status data along with header and other information. A pointer is available to indicate the location of the last frame received. When last setpoint and command data is read, the data is stored in separate registers. The SNS will operate at 60 Hz. If data is read on each pulse, then the memory will hold the last 90 seconds of data. When a beam dump occurs memory updating can be inhibited which preserves a data history up to 90 seconds prior to the event.

The PSC keeps track of link errors. For each channel there are bits for link error, checksum error or channel disabled. When these errors are received they are latched until cleared by the operator. There are commands to disable or enable any channel.

The PSC supports burst mode that causes the PSC to poll the PSI at a fixed rate for a number of cycles with the data being stored in the circular buffer. The maximum rate is 10,000 Hz. This can be used to look for unusual conditions or measure the power supply response time.

## 1.4 Test Setup

Figure2 is the configuration of a lab setup used to test the hardware under high load conditions. There is an IOC with 8 PSCs with 24 PSIs connected to four of the PSCs. The PSCs are using external triggers to read and write to the PSIs. The trigger rate is set at 4000Hz in this example to test operation under high load. For each PSI, the output voltage is fed back to the input connectors and the output commands are fed back to the status input to simulate a power supply.

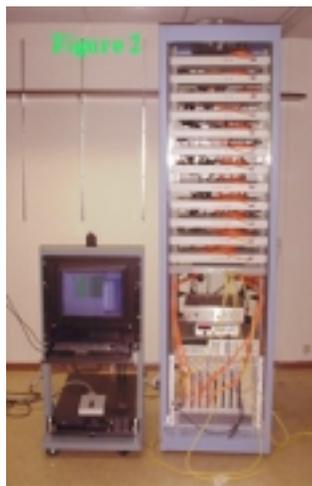

## 2 SOFTWARE

### 2.1 Serial Port Software

The PSC has a serial port that allows communicating with power supplies using just a single PSC/PSI pair and a laptop computer.

Figure 3 is a test setup showing the standalone PSC connected to a PSI with Labview software running on a laptop. This system is made available to hardware manufacturers to check that their equipment is compatible with the control system.

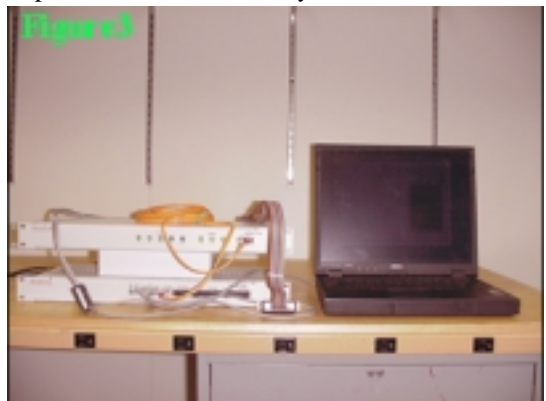

Figure 3 – Standalone PSC Configuration

A Labview library is now available that supports all functions of the PSC/PSI via a simple serial port. Labview has been used extensively to test the PSI.

Using the serial port, all functions of the PSC can be completely exercised. Setpoints and commands can be issued, data can be read, burst mode parameters can be set, burst mode initiated, communication status can be monitored, errors detected etc.  The serial port is limited to reading or writing at about 15Hz. For fast reads, burst mode is used and the data is stored in memory. Data in memory can be read at a slow rate.

Users can easily build their own applications to handle up to 6 power supplies based on the library. Many VIs were built based on this library to checkout the PSC and PSI hardware. For instance we checked accuracy and linearity of DAC and ADCs and temperature stability of PSI. Software is available that will measure the response and accuracy of DAC and ADC for the full range of input and output voltages.

VIs were written to test power supplies. They are being used to do prototype testing of power supplies from three manufacturers: one in Canada, one in California and another in Denmark. When the units are shipped from the manufacturer, the control hardware (PSI) will already be integrated into and tested with the power supply hardware. This reduces installation and commissioning time.

If you combine this library with ActiveX channel access server, you can interface with an EPICS application.

### 2.2 Epics Driver and Device Interface

For the SNS project, our control platform is EPICS. A general vxWorks driver and EPICS support interface were built for the PSC. All functions of the hardware are supported.

Major functions are listed below:

- Initialize PSC
- Read digital status and analog input

- Read history waveform
- Save history memory to file to analyze
- Send setpoint and command to power supply
- Set burst mode rates and length
- Get communication status
- Tolerance check and command-readback consistency check
- Set PSC operation mode: normal or burst
- Enable/disable channel to PSI
- Choose software trigger or hardware trigger
- Set unipolar or bipolar mode
- Average analog readback to increase resolution

The software is now being used to do high data rate system tests. Our record scan rate can be more than 80Hz and read data rates of several hundred Hz.

We can also run the software in a portable system. Figure 4 is the portable configuration for VME based test. It consists of a small VME crate with computer and PSC connected to a remote PSI. The laptop loads the computer with an Epics application and runs MEDM display software.

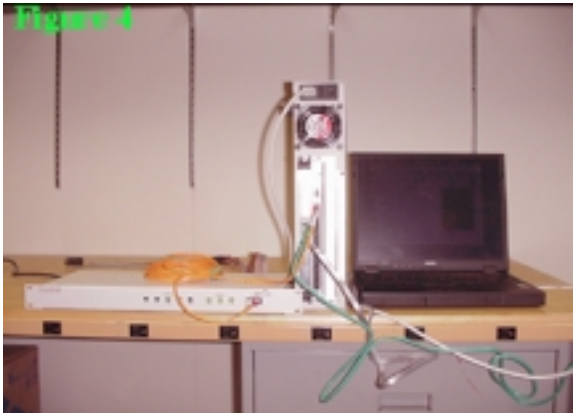
Figure 4 – Portable EPICS Configuration

Because the driver itself is not EPICS-dependent, the user can easily integrate PSC/PSIs with any vxWorks based system, not only EPICS.

## 2.3 MEDM and high level tools

MEDM screens have been developed and are being used to check operation of the hardware. In the future, we will develop workstation tools to read data from PSC memory, save to disk, and analyze the power supply data for ripple, glitches etc.

## 3 TEST RESULTS

PSC and PSI hardware has been delivered to BNL and is undergoing extensive testing. Two configurations are being tested. One system with 24 PSIs connected uses external hardware triggers to read the PSI at rates up to 4000 Hz. This system was tested for over a week without any errors noted. The second is similar to the first except that it uses software to trigger the reads. This system is undergoing tests.

## 4 STATUS

The PSC/PSI combination will also be used by the BAF project at BNL and there are plans to use it in future additions or upgrades in the CA department. BNL, ORNL and LANL will be buying power supplies compatible with this interface. Several manufacturers will be providing supplies to SNS compatible with this interface.

## 5 SUMMARY

The standard power supply interface is expected to greatly simplify the installation of power supplies at ORNL. The power supply vendors will test the power supplies at the factory with PSC/PSI hardware. Labview software is provided that allows complete power supply testing. When the power supplies arrive at the site they will already have been tested with control system hardware. At the site it will only be necessary to connect a pair of fiber cables between the control system and PSI to have an operational system.